# Time localization of energy in disordered time-modulated systems

Benjamin Apffel[1,2], Sander Wildeman[1,2], Antonin Eddi[2] & Emmanuel Fort[1,*]

[1] ESPCI Paris, PSL University, CNRS, Université de Paris, Institut Langevin, 1 rue Jussieu, F-75005 Paris, France.

[2] PMMH, CNRS, ESPCI Paris, Université PSL, Sorbonne Université, Université de Paris, F-75005, Paris, France.

*Corresponding author: emmanuel.fort@espci.fr

**Abstract**

Wave localization induced by spatial disorder is ubiquitous in physics. Here, we study the temporal analog of such phenomenon on water waves. Our time disordered media consists in a collection of temporal interfaces achieved through electrostriction between water surface and an electrode. The wave field observed is the result of the interferences between reflected and refracted waves on the interfaces. Although no eigenmode can be associated to the wave field, several common features between space and time emerge. The waves grow exponentially depending on the noise level in agreement with a 2D matrix evolution model such as in the spatial case. The relative position of the momentum-gap appearing in the time modulated systems plays a central role in the wave field evolution. When tuning the excitation to compensate for the damping, transient waves, localized in time, appear on the liquid surface. They result from a particular history of the multiples interferences produced by a specific sequence of time boundaries.

The localization and focusing of the waves induced by disorder in the propagation medium is observed in many contexts[1]. Examples can be found in quantum[2–4], optical[5,6], acoustic[7], seismic[8] or hydrodynamic[9–11] systems. When a wave travels through a disordered medium interferences between counter-propagating waves result in enhanced backscattering and weak localization[5,12]. Upon increasing the disorder, waves propagation is inhibited and undergoes Anderson localization[2,6]. In weak disorder landscapes whose fluctuations are correlated on length scales greater than the wavelength, the waves form caustics and branched transport channels with strongly enhanced intensity[9,13,14].

Because time and space are, to some extent, interchangeable in wave propagation, several phenomena observed in spatially modulated systems have a temporal counterpart in the time modulated ones[15–21]. A type of temporal localization, called dynamical localization, have been previously studied in the dual momentum space with "kicked rotor" systems[22] or with quantum particles confined spatially in a time varying potential through a theoretical approach[23,24]. Here, we study the case of spatially invariant system for a direct exchanged between space and time variables. What are the consequences of time disorder? Does energy localization occur in the time domain? First, we present a model based on a spatially invariant system submitted to a Dirac comb excitation with controlled noise. We then show experimental results obtained on electrically excited water waves and discuss the various effects of temporal noise on the energy localization in time.

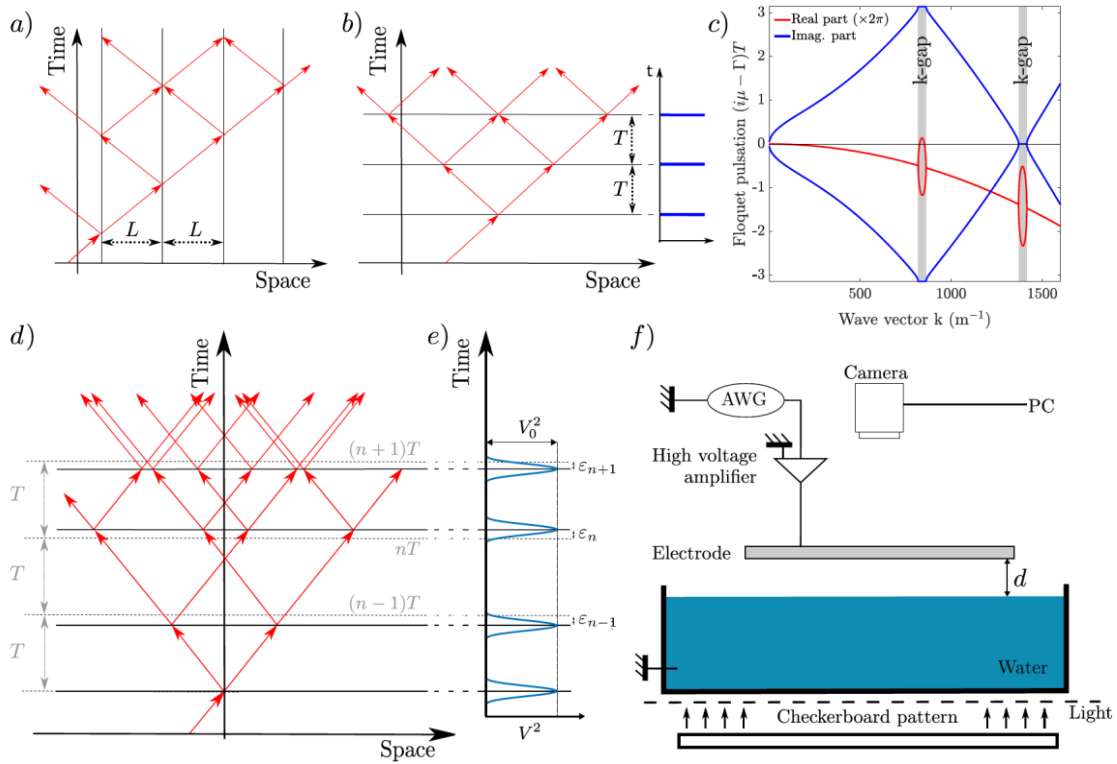

**Figure 1**: Schematics of the wave front reflections and transmissions on a spatial Dirac comb (a) and a temporal one (b). (c) Floquet diagram with vertical $k$-gaps for a temporal Dirac comb with damping ($V_0 = 8.0$ kV, $d = 5$ mm and $2\nu = 7 \cdot 10^{-6}$ m²/s). (d) Schematics of the wave front propagation in a temporal Dirac comb with noise obtained by shifting the time interfaces and (e) associated electric forcing $V(t)^2$. (f) Schematics of the experimental setup with the amplitude wave generator (AWG) controlling the voltage of the electrode.

Our model is based on a spatially invariant system submitted to a temporal modulation in the shape of a Dirac comb with a controlled level of disorder. The Dirac comb is a standard model used for spatially periodic potentials in which a modulated potential is replaced with discrete boundary conditions imposed on the wave function[25]. Figure 1a shows a schematics of a wave front propagating in a 1D Dirac comb potential with the multiple reflections and transmissions at each interface. Figure 1b shows the time analog with the Dirac comb composed of time interfaces induced by the periodic forcing. The time reflections and transmissions on the interfaces generate counter-propagating and co-propagating waves respectively[26,27]. In both cases, the wave field is the result of the interferences of these multiple waves generated at the interfaces.

We use liquids to study the effect of disorder in time modulated systems. Large amplitude and versatile time modulations can be easily implemented in liquids[28–30]. A time interface has been obtained for instance by applying a sudden vertical jolt. An initial wave propagating on the liquid produced a co-propagating and a counter-propagating wave which are respectively associated to the transmitted and the reflected wave at the time interface[26]. Here, we use an alternative versatile excitation based on electrostriction using a flat electrode placed at a distance $d$ above the grounded conductive water surface[31–33]. The electric field exerts an attractive force on the liquid surface which modifies the wave speed $c(k)$ for a wavenumber $k$ satisfying

$$c(k)^2 = c_0^2(1 - \alpha(t)) \quad \text{with } \alpha(t) = \chi_0 V(t)^2 \text{ and } \chi_0 = \epsilon/(\rho c_0^2 d^2 \tanh(kd)) \quad (1)$$

$V(t)$ is the electric potential, $c_0$ the wave speed for $V(t) = 0$ given by the gravity-capillary dispersion relation, $\epsilon$ the dielectric permittivity of the air and $\rho$ the density of the liquid[31–33]. Electric pulses of maximum amplitude $V_0$ and duration $\tau_p$ at successive discrete times $T_n$, $n$ being an integer, are modelled as a Dirac comb $V(t)^2 = fV_0^2\tau_p \sum_{n\geq 0} \delta(t - T_n)$ with $f$ accounting for the shape of the pulse. In the experiments, $\chi_0 V_0^2$ is typically of the order of 0.3-0.5.

The time evolution of the wave field $\phi(\mathbf{k}, t)$ of wavevector $\mathbf{k}$ is given by the following non homogeneous wave equation[26]

$$\frac{\partial^2 \phi}{\partial t^2}(\mathbf{k}, t) + 2\Gamma \frac{\partial \phi}{\partial t}(\mathbf{k}, t) + \omega_0^2 \phi(\mathbf{k}, t) = \omega_0^2 \beta_0 \sum_{n\geq 0} \delta(t - T_n) \phi(\mathbf{k}, t) \quad (2)$$

with $\omega_0 = c_0 k$ the wave angular frequency, $\beta_0 = f\chi_0 V_0^2 \tau_p$ and $\Gamma$ the wave damping rate due to viscosity, $\Gamma = 2\nu k^2$ with $\nu$ bring the kinematic viscosity of the liquid[34]. In practice, $1/\Gamma \gg \tau_p$. The time interfaces can be interpreted as sources proportional to the wave field $\phi(\mathbf{k}, T_n)$ at the time of the electric pulse[26]. We solve eq (2) using a matrix transfer approach. The evolution of the wave field is completely characterized by $\Psi = \left[\phi \ \frac{1}{\omega_0}\frac{\partial \phi}{\partial t}\right]^T$. The crossing of the time interface is given by $K = \begin{bmatrix} 1 & 0 \\ \omega_0 \beta_0 & 1 \end{bmatrix}$ and the propagation during $\Delta T_n = T_{n+1} - T_n$ between two successive time interfaces is given by $R(\Delta T_n) = e^{-\Gamma \Delta T_n} \begin{bmatrix} \cos(\omega_0 \Delta T_n) & \sin(\omega_0 \Delta T_n) \\ -\sin(\omega_0 \Delta T_n) & \cos(\omega_0 \Delta T_n) \end{bmatrix}$.

Thus, the evolution of the wave field satisfies $\Psi_{n+1} = M_n \Psi_n = R(\Delta T_n) K \Psi_n$ which by recurrence gives $\Psi_{n+1} = \left(\prod_{p\leq n} M_p\right) \Psi_0$. For a periodic excitation, $\Delta T_n = T$, $M_0 = R(T)K$ and $\Psi_{n+1} = M_0^{n+1} \Psi_0$. Following Floquet analysis, the two eigenvalues of $M_0$ can be written $\lambda_j = e^{(i\mu_j - \Gamma)T}$ with $\mu_j$ a complex value and $j = 1$ or $2$. Figure 1c shows the real part of $(i\mu_j - \Gamma)T$ (blue lines) and its imaginary part (red lines) as a function of $k$. Vertical momentum $k$-gaps associated with real values of $\mu_j$ are clearly visible. They are the analog of energy-gaps in spatial crystals. However, while the latter are forbidden admitting only exponentially decaying solutions due to energy conservation, $k$-gaps also allow exponentially increasing solutions, as energy can be supplied by the forcing. For $\Re(i\mu_j - \Gamma) > 0$, these solutions lead to the Faraday instability at half the excitation frequency[35].

We now introduce noise in the system as a random time shift such that $T_n = (n + \epsilon_n)T$, $\epsilon_n$ being a variable taken independently and uniformly in $[-\sqrt{3}\sigma, \sqrt{3}\sigma]$ with $\sigma$ the noise standard deviation (see Fig. 1d). The evolution matrices $M_n$ are now random and correlated depending on $\epsilon_{n+1} - \epsilon_n$. However, they can be redefined uncorrelated by setting $\Psi'_n = R(-\epsilon_n T)\Psi_n$ and

$M'_n = R(-\epsilon_n T)M_n R(\epsilon_n T)$ to apply the Fürstenberg theorem[36] which states that $\Psi'_n \simeq \exp(\upsilon n)$ for $n$ large enough, with $\upsilon$ being the Lyapunov exponent. This should lead to a localization behavior similar to the one observed in disordered spatial crystals[37].

The experimental setup is shown in Figure 1e. It consists of a 30x30x3 cm³ plexiglass container filled with tap water. A transparent ITO electrode deposited on a glass is suspended horizontally at a distance $d = 5$ mm over the electrically grounded water. The electric potential $V(t)$ consists of narrow peaks of amplitudes $V_0$ in the range of 6 to 8 kV with a repetition rate of $\omega_0/2\pi = 60$ Hz. The pulses are arch of sinus of duration $\tau_p = 0.4T$ and maximum amplitude $V_0$ (see Fig. 1e). Matching the integrals $V(t)^2$ to fit the Dirac comb model gives $\beta_0 = (3\tau_p/8)\chi_0 V_0^2$. An electrically induced Faraday instability is triggered above a certain $V_0^2$ threshold (~ 8 kV) with waves oscillating at half the forcing frequency $\omega_0/4\pi$=30 Hz. The wave field is measured from images taken at 90 fps with a camera mounted above the bath using the deformation of a checkerboard pattern placed below[38]. The amplitude of the Faraday waves $A_F(t)$ at time $t$ is obtained by applying a time filter at $\omega_0/4\pi$ and spatial averaging under the electrode.

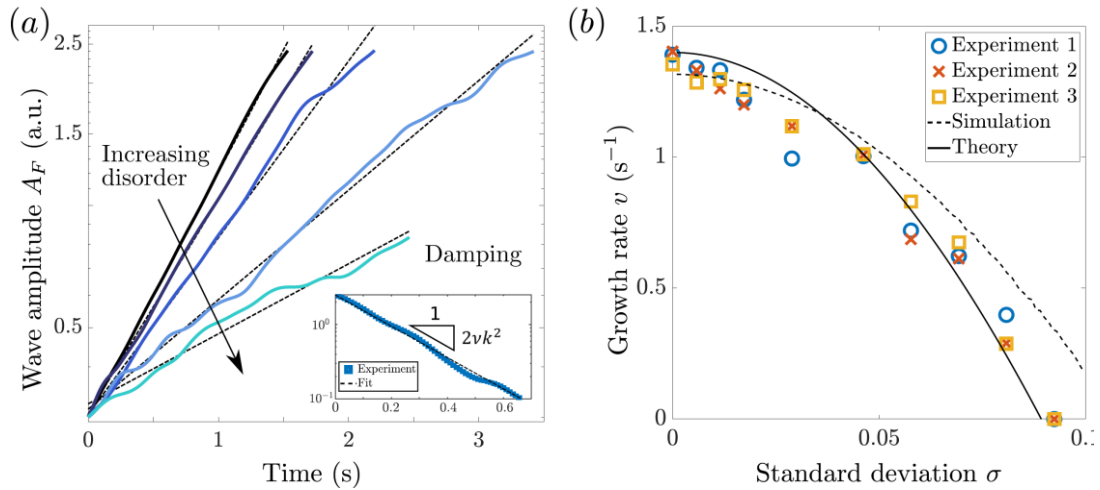

**Figure 2**: (d) Growth of the Faraday wave amplitude $A_F(t)$ with time for various noise levels $\sigma = 0, 0.017, 0.046, 0.069$ and $0.081$. The Lyapunov exponents $\upsilon$ are obtained from the exponential fits (dashed lines). Inset: Decay of $A_F(t)$ with time when the forcing is turned off. The damping rate $\Gamma$ is obtained from an exponential fit (dashed line). (b) Fitted Lyapunov exponents $\upsilon$ as a function of $\sigma$ for various experiments. Numerical simulations from the matrix model (dotted line) taking into account the fitted $\Gamma$ and theoretical model (full line) based on the decrease of the forcing component at $\omega_0$.

We first focus on the effect of disorder on the exponential growth of the wave. Figure 2a shows a typical example of the time evolution of $A_F(t)$ for various noise standard deviations $\sigma$ from $\sigma = 0$ to $8.1 \cdot 10^{-2}$. The amplitude of the electric pulses $V_0$ is set to 7 kV to trigger the Faraday instability. In agreement with the Fürstenberg theorem[36], the wave grows exponentially for small enough amplitudes ($A_F(t) \ll \lambda_F$) for which non-linear hydrodynamic effects are negligible. The fitted Lyapunov exponents $\upsilon$ (dashed lines) decrease with increasing noise. A characteristic damping time $1/\Gamma = 0.19$ s is measured from the decay of the Faraday waves when excitation is stopped (see inset Fig. 2a). This value used in the matrix model yields to the Faraday threshold value at $V_0 = 7$ kV which agrees with experiments. From the value of $\Gamma$ we can extract $2\nu = 7.10^{-6}$ m²/s that is higher than the expected value for water but of the right order of magnitude[21].

Figure 2b shows the fitted exponents $\upsilon$ as a function of the noise level for a series of experiments with different noise sequences. Each sequence is run three times to ensure that the measured growth rates are robust to experimental drifts. The decrease of $\upsilon$ with increasing noise level $\sigma$ can be accurately reproduced by the numerical calculations with the matrix model using the fitted damping rate $\Gamma$ (see dashed line Fig. 2b). It can be ascribed to a growing random phase in the interfering reflections and transmissions on the time interfaces which reduces the wave field amplitude of the otherwise perfectly in phase multiple contributions. This result can also be interpreted in the spectral domain. Faraday instability is directly related to the oscillation of $V(t)^2$ at $\omega_0$. For small noise levels, $\upsilon$ is expected to follow $\upsilon(\sigma) \propto (\widehat{V^2}(\sigma) - \widehat{V_F^2})$, with $\widehat{V^2}(\sigma)$ and $\widehat{V_F^2}$ being the Fourier components of $V^2(t)$ at $\omega_0$ in the presence of noise and when forcing is set at the Faraday threshold respectively[39]. From the central limit theorem, one can show that $\widehat{V^2}(\sigma) = \widehat{V^2}(0)\text{sinc}(2\pi\sqrt{3}\sigma) \approx \widehat{V^2}(0)(1 - 2\pi^2\sigma^2)$ for small $\sigma$ and a large number of pulses. The resulting quadratic shape $\upsilon(\sigma) = \upsilon(0)(1 - \zeta\sigma^2)$ is in good agreement with experimental findings and can be adjusted by setting $\upsilon(0) \approx 1.4$ and $\zeta \approx 120$ (see Fig. 2b solid line).

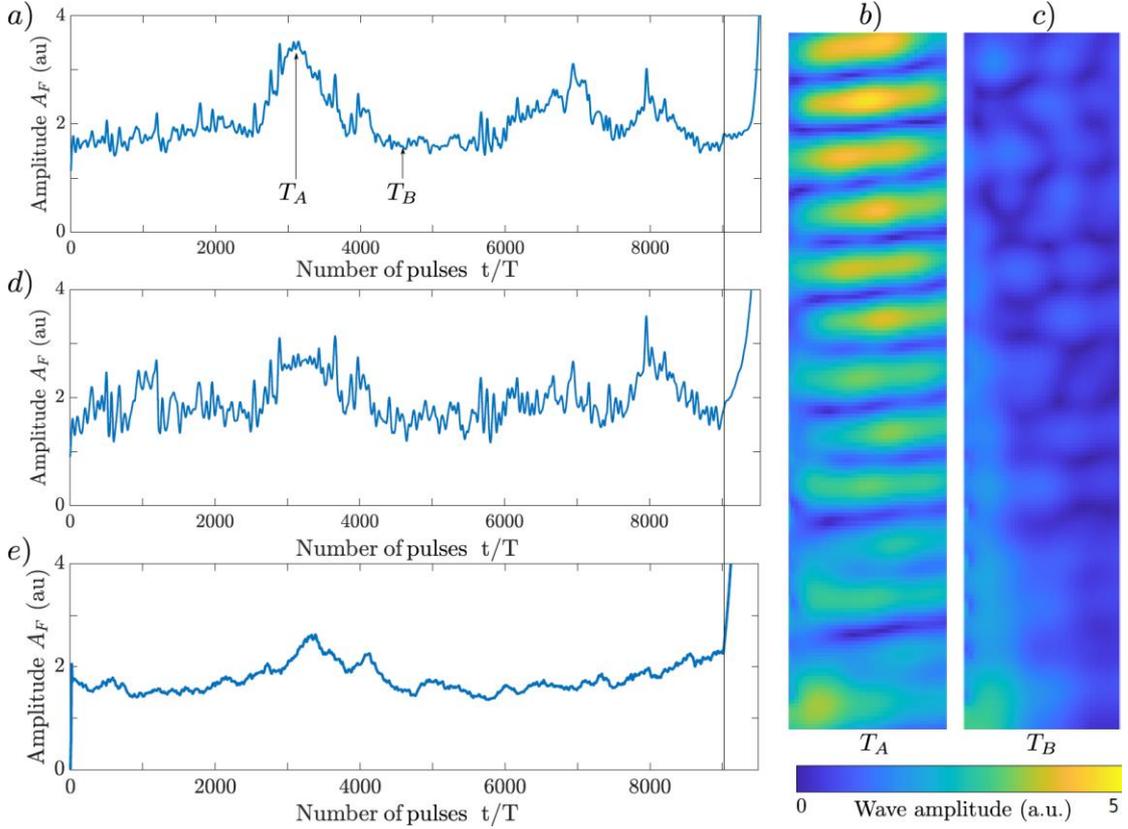

**Figure 3**: (a) Time evolution of the Faraday wave amplitude $A_F(t)$ for a periodic excitation with a given noise sequence with $\sigma = 0.1/\sqrt{3}$. $V_0$ is tuned at the Faraday instability threshold. The noise is suppressed at $t > 160$ s (vertical line). (b) and (c) Wave fields at the Faraday frequency at time $T_A$ and $T_B$ respectively. (d) $A_F(t)$ for the same excitation sequence as in (a). (e) Simulated time evolution of $A_F(t)$ using the matrix model for the same excitation sequence as in (a) with $V_0 = 7.322$ kV, $\nu = 7 \cdot 10^{-6}$ m²/s and $d = 5$ mm.

We now focus on the amplitude fluctuations of the wave field, which are known to contain significant information on localization processes[40]. Temporal fluctuations of the wave field can be observed by tuning $V_0$ at the Faraday threshold for a chosen noise level $\sigma$ to achieve a null Lyapunov exponent. The average energy gain induced by the random time interfaces thus compensates for the damping. Figure 3a shows the amplitude of the Faraday waves $A_F(t)$ as a function of time $t/T$ for a given pulse sequence. The periodic excitation with added noise last 160 seconds. Then, the noise is removed (vertical line) from the periodic excitation to measure the exponential growth of the wave amplitude. Localized peaks are observed in the Faraday waves with typical durations extending over hundreds of periods $T$. Figures 3b and 3c show the Faraday wave field during and outside a peak at time $T_A$ and $T_B$ respectively. When the same excitation sequence is applied again, the measured wave fluctuations (see Fig. 3d) are highly correlated in both experiments. This clearly show that these fluctuations are the result of a specific temporal sequence and that the multi-wave interference process which produce them is experimentally reproducible. Figure 3e shows the simulated fluctuations with the same sequence using the matrix model. Although the correlation is lower (0.55), the central double-peak feature is still visible.

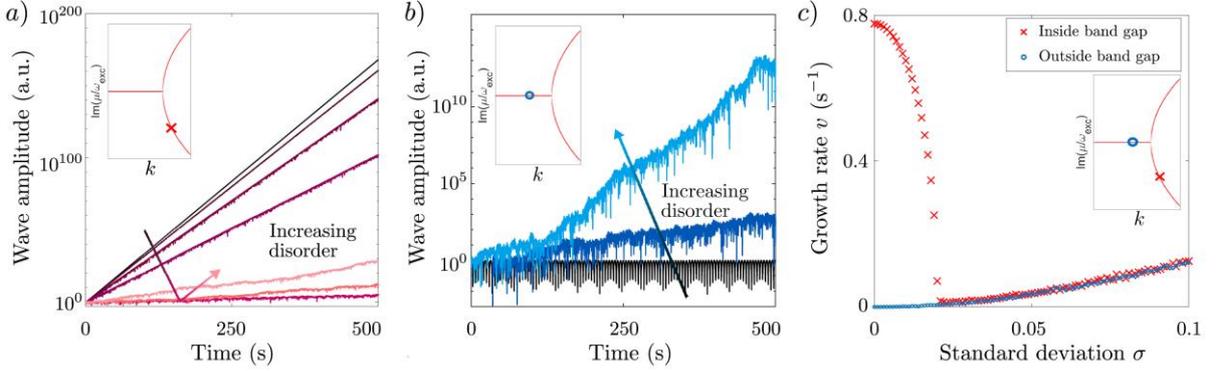

**Figure 4**: Simulation of the wave growth with time for different noise standard deviations for waves (a) just inside and (b) outside the $k$-gap with wavenumber $k_{in} = 819.5$ m$^{-1}$ and $k_{out} = 819.2$ m$^{-1}$ respectively (see insets) and $\sigma = 0.1, 0.6, 1.1, 1.6, 2.1, 7, 10 \cdot 10^{-2}$ and $\sigma = 0.1, 3, 6 \cdot 10^{-2}$ respectively. Linear fits (not shown) give the Lyapunov exponents $v$. (c) Fitted exponents $v$ as a function of $\sigma$.

The disorder also changes significantly the waves other than the most unstable Faraday mode. In the experiments, observations of these waves is hindered by the presence of the larger amplitude Faraday waves. However, numerical simulations enable their study by setting $\Gamma = 0$, $V_0 = 8.0$ kV and d = 5 mm and changing the noise level $\sigma$. We study two modes at the edge of the $k$-gap, with a wavenumber $k_{in}$ just inside ($k_{in} = 819.5$ m$^{-1}$) and $k_{out}$ outside ($k_{out} = 819.2$ m$^{-1}$) the gap respectively. As shown in Figure 4a, the wave amplitude at $k_{in}$ (see inset) grows exponentially in all cases for various disorder levels in agreement with the model. As the noise level increases, the fitted exponents $v$ first decrease at low noise levels ($0 < \sigma < 0.025$) and then increase at higher noise levels ($\sigma > 0.025$). Outside the $k$-gap, the growth is also exponential but in this case the exponent increases monotonously with increasing noise on the whole $\sigma$ range, $0 < \sigma < 0.1$. The evolution of the exponents $v$ with $\sigma$ are plotted in Fig. 4c for the two cases $k_{in}$ (red crosses) and $k_{out}$ (blue circles). The two curves superimpose in the regions where $v$ increases with $\sigma$. The decreasing behavior of $v$ at low noise level for $k_{in}$ is similar to the experimental observations obtained at the Faraday frequency (see Fig. 2a and

2b). These results are consistent with an effective shrinking of the band gap as the periodicity deteriorates with the noise. . The validity of the Fürstenberg theorem implies that the waves must grow exponentially as observed either inside or outside the gap for arbitrary noise levels.

Contrary to spatial interfaces, the crossing of a time interface is non-unitary meaning that the energy of the wave field is not conserved across the boundary. From the expression of the matrix $K$, in the case of a monochromatic wave at $\omega_0$, an incident wave field defined by $\Psi^- = \left[\phi^- \; \frac{1}{\omega_0}\frac{\partial \phi^-}{\partial t}\right]^T$ produces an additional wave field $i\omega_0\beta_0 \left[0 \; \frac{1}{\omega_0}\frac{\partial \phi^-}{\partial t}\right]^T$ when crossing the interface (using $\phi^- = i\frac{1}{\omega_0}\frac{\partial \phi^-}{\partial t}$). Using the superposition principle, this can be written as the sum of a forward propagating wave $\frac{i\omega_0\beta_0}{2}\left[\phi^- \; \frac{1}{\omega_0}\frac{\partial \phi^-}{\partial t}\right]^T$ and a time-reversed backward propagating one $-\frac{i\omega_0\beta_0}{2}\left[\phi^- \; -\frac{1}{\omega_0}\frac{\partial \phi^-}{\partial t}\right]^T$. This interpretation highlights the momentum conservation with the symmetric production of counter-propagative waves as well as the non-conservation of energy with the creation of waves. For an incident wave, the time interface can be characterized by a transmission coefficient $t = 1 + \frac{i\omega_0\beta_0}{2}$ and the reflection one $r = -\frac{i\omega_0\beta_0}{2}$. The general expression for the incident wave is two counter-propagating waves $\phi^-(\mathbf{k},t) = Ae^{i\mathbf{k}.\mathbf{r}+i\omega_0 t} + Be^{i\mathbf{k}.\mathbf{r}-i\omega_0 t}$ with the total energy $E^- \propto |A|^2 + |B|^2$, $A$ and $B$ being complex values. Just after the boundary, the field becomes $\phi^+(\mathbf{k}) = [tA + r^*B]e^{i\mathbf{k}.\mathbf{r}+i\omega_0 t} + [rA + t^*B]e^{i\mathbf{k}.\mathbf{r}-i\omega_0 t}$. For an incident propagating wave ($B = 0$), the time interface creates a standing wave with limited amplitude in second order in $\omega_0\beta_0$ ($\Delta E = \frac{\omega_0^2\beta_0^2}{2}E^-$). For an incident standing wave, the interface also generates a standing wave with an associated energy $\Delta E \approx (\omega_0\beta_0 \sin\varphi)E^-$ with $B = Ae^{-i\varphi}$. It yields to an energy increase or decay depending on the interference with the incident wave field. To accumulate energy in the time crystal, the phase condition $\varphi$ must also be recovered after the propagation between two successive interfaces. The relative phase between the two counter propagative waves must thus change by $2\pi n$, $n$ being an integer between two interfaces. This leads to frequencies $n\omega_0/2$ in the $k$-gaps. The maximum energy output corresponds to the Faraday instability at $\varphi \approx \pi/2$. When moving to the $k$-gap limits, smaller energy gains are achieved when crossing the time interfaces, reaching zero at the $k$-gap limit (for $\varphi \approx 0$ or $\pi$). Note that an associated symmetric over-damped mode exists for the opposite phase $\varphi$.[21] The presence of a random time shift between the interfaces results in fluctuations of the phase-lock condition which in turn changes the energy gain at each interface. For the Faraday mode, since $\Delta E$ is maximal, any perturbation results in a decrease of wave growth as observed experimentally (see Fig. 2b). For modes such as $k_{in}$ for which is $\Delta E$ not maximal, the output is more complex to infer due to the possible gain or loss of energy resulting from the interplay the perturbed phase shift acquired between two interfaces and the amplitude of $\Delta E$ at the crossing of each interface[41].

In summary, time localization induced by noise differ fundamentally from its spatial counterpart since it is not related possible to describe these localizations in terms of mode localization related to an Hamiltonian eigenstate. This is a consequence of the non-unitarily of the evolution operator and the non-conservation of energy along time. However, several features are common. Time disorder induces a temporal localization of the energy with features that are similar to its spatial analog. An interesting perspective of these results would be to generalize the effect of noise on wave propagation with a spatio-temporal disorder and to study the spatio-temporal localization of energy. In addition, this could be implemented experimentally with the use of several electrodes driven independently to realize time-varying inhomogeneous energy landscapes.

**Acknowledgments**

The authors would like to thank Rémi Carminati for fruitful discussions. We thank the support of AXA Research Fund and the French National Research Agency LABEX WIFI (ANR-10-LABX-24).